\documentclass[aps,onecolumn,a4paper,oneside,pra,
nobibnotes,nofootinbib,amsfonts,amssymb,amsmath,showpacs,
showkeys,11pt,tightenlines]{revtex4}
\usepackage{amssymb,amsmath}
\usepackage{hyperref}
\usepackage{graphicx}
\usepackage{subfigure}
\begin{document}
\title{\bf Quasi-Hermitian Hamiltonians associated with exceptional orthogonal polynomials }
\author{Bikashkali Midya \footnote{Fax: ~ +91 3325773026} }
\email{bikash.midya@gmail.com}
\affiliation{Physics and Applied Mathematics Unit, Indian Statistical
 Institute,\\ Kolkata 700108, India.}
\vspace{2 cm}

\begin{abstract}
{\bf Abstract:}

 Using the method of point canonical transformation, we derive some exactly solvable rationally extended
quantum Hamiltonians which are non-Hermitian in nature and whose bound state wave functions are associated with Laguerre or Jacobi-type $X_1$ exceptional orthogonal polynomials.
These Hamiltonians are shown, with the help of imaginary shift of co-ordinate: $ e^{-\alpha p} x e^{\alpha p} = x+ i  \alpha $,
to be both quasi and pseudo-Hermitian. It turns out that the corresponding energy spectra is entirely real.
\end{abstract}


\maketitle

\vspace{1.25 cm}
\noindent
\section{Introduction}
Since the discovery \cite{UKM09,UKM10a}  of $X_1$ exceptional orthogonal polynomials (here after EOPs) in mathematical physics there has been renewed interest in
the analysis of exactly solvable shape invariant quantum systems.
Unlike the classical orthogonal polynomials, these new polynomials have the remarkable properties \cite{UKM10b} that they
still form complete sets with respect to some positive definite measure, although they start with degree $n \ge 1$ polynomials instead of a constant.
Laguerre and Jacobi-type $X_1$ EOPs have made their appearance in the bound state wave functions of the quantum systems with both constant \cite{Qu08,Qu09} and position-dependent
 mass \cite{MR09}. These quantum systems are shown \cite{BQR09},
 with the help of reducible second order supersymmetric transformation, to be rationally extended version of conventional ones associated with the classical orthogonal polynomials.
 This supersymmetric transformation also explains the isospectrality of the conventional and rationally extended potentials.
 Subsequently, EOPs are generalized to higher co-dimension \cite{OS09a,OS10,UKM12} and to multi-indexed systems \cite{OS11b,UKM12b} and
associated shape invariant Hamiltonians are reported
\cite{Gr11a,Gr11b,STZ10,Qu11,Qu11a}. Some properties of these polynomials are studied in ref. \cite{HOS11}. EOPs are also used in connection with
discrete quantum mechanics \cite{OS09b,OS11a}, Dirac and Fokker-Planck equations \cite{Ho11b},
 pre-potential approach \cite{Ho11a}, information entropy \cite{DR11}, quantum Hamilton Jacobi formalism \cite{RP+12},
 dynamical breaking of higher order supersymmetry \cite{MRT12} and quasi-exactly solvable problems \cite{Ta10}. However, the application of these new polynomials to
 the non-Hermitian quantum systems is not reported so far.\\

Non-Hermitian Hamiltonians are important due to the fact that, despite being non-Hermitian in nature,
these operators may constitute unitary quantum mechanical systems \cite{SGH92,Be07,Mo10}. Non-Hermitian parity-time (${\cal{PT}}$) symmetric Hamiltonians
possess real discrete energy eigenvalues if the corresponding eigenfunctions are also ${\cal{PT}}$ symmetric, otherwise the eigenvalues
occur in complex conjugate pairs \cite{BB98}.
$\mathcal{PT}$-symmetric Hamiltonian having all eigenvalues real is connected to the existence of a positive definite inner product
$ \langle\psi,\eta \phi\rangle$ which render the Hamiltonian $H$ to be pseudo-Hermitian  \cite{Mo02a,Mo02b,Mo02c,DG09} $H^\dag = \eta H \eta^{-1}$, where the Hermitian linear automorphism $\eta : {\cal{H}} \rightarrow {\cal{H}}$ is bounded and positive definite.
Another equivalent condition for the reality of the energy spectrum $H$ is the quasi-Hermiticity \cite{KS04,SGH92,ZG06,MB04}, i.e.
the existence of a invertible operator $\rho$ such that $h = \rho H \rho^{-1}$ is Hermitian with
respect to usual inner product $\langle \psi| \phi\rangle$. Quasi-Hermitian Hamiltonian shares the
same energy spectrum of the equivalent Hermitian Hamiltonian $h$ and the wave functions are obtained by operating $\rho^{-1}$
on those of $h$. Most of the analytically solvable non-Hermitian Hamiltonians are constructed by making the coupling
constant of the known exactly solvable potentials imaginary \cite{Ah01b,BR00,BQ02,Le}. In some other cases \cite{LZ00,LZ01,Zn03},
 the coordinate is shifted
with an imaginary constant. Several of these classes of Hamiltonians are argued to be pseudo-Hermitian under $\eta = e^{-\alpha p}$ \cite{Ah01,Ah02}.
 For a real $\alpha$ and $p=-i\frac{d}{dx}$, the operator $\eta$ shifts the coordinate $x$ to $x + i\alpha$.\\

The goal of this letter is to generate some rationally extended Hamiltonians which are non-Hermitian in nature
and whose bound state solutions are associated with Laguerre or Jacobi-type $X_1$ exceptional orthogonal polynomials. By `rationally extended Hamiltonians'
we mean those which are the extensions of the well known Hamiltonians by addition of some rational functions. The method of point canonical transformation (PCT) \cite{BS62,Le89}, which consists of
 transformation of the initial Schr\"{o}dinger equation to a differential equation of some special function, has been used here to achieve our goal.
The non-Hermiticity enters, in a natural way, into the potentials through the purely imaginary
constant of integration appears in PCT. We also show, with the help of a similarity transformation, that the new non-Hermitian
 Hamiltonians obtained here are quasi as well as pseudo-Hermitian.  In particular, it has been identified that the
 positive definite operators $\rho = e^{-\frac{\alpha}{2} p}$ and $\eta = \rho^2$ play the roles of a quasi and pseudo-Hermitian operators respectively.\\

 \section{Quasi-Hermitian Hamiltonians associated with Laguerre or Jacobi type $X_1$ EOPs}
 \noindent
 Here we use the method of PCT to derive some exactly solvable non-Hermitian Hamiltonians whose bound state wave functions are associated with Laguerre or Jacobi type $X_1$
 exceptional orthogonal polynomials. For this we first briefly recall the method of point canonical transformation.

 In PCT approach \cite{BS62,Le89}, the general solution of the Schr\"{o}dinger equation (with $\hslash = 2 m = 1$)
\begin{equation}
 H \psi(x) = -\frac{d^2\psi(x)}{dx^2} + V(x) \psi(x) = E \psi(x)\label{e1}
\end{equation}
can be assumed as
\begin{equation}
 \psi(x) \sim f(x) F(g(x)) \label{e18}
\end{equation}
where $F(g)$ satisfies the second order linear differential equation of a special function
\begin{equation}
 \frac{d^2 F}{dg^2} + Q(g) \frac{dF}{dg} + R(g) F(g) =0.\label{e2}
\end{equation}
Substituting the assumed solution $\psi(x)$ in equation (\ref{e1}) and comparing the resulting equation with the equation (\ref{e2})
one obtains the following two equations for $Q(g(x))$ and $R(g(x))$
\begin{subequations}
  \begin{equation}
    Q(g) = \frac{g''}{g'^2} + \frac{2 f'}{f g'}
  \end{equation}
  \begin{equation}
    R(g) = \frac{E-V(x)}{g'^2} + \frac{f''}{f g'^2}
  \end{equation}
\end{subequations}
respectively. After some algebraic manipulations above two equations reduces to
\begin{subequations}
\begin{equation}
 f(x) \approx g'(x)^{-1/2} ~ e^{^{\frac{1}{2} \int Q(g) dg}},\label{e7}
\end{equation}
\begin{equation}
 E-V(x) = \frac{g'''}{2g'} - \frac{3}{4} \left(\frac{g''}{g'}\right)^2 + g'^2 \left(R-\frac{1}{2} \frac{dQ}{dg}
 -\frac{1}{4} Q^2\right).\label{e3}
\end{equation}
\end{subequations}
Now, we are in a position to choose the special function $F(g)$ (consequently $Q(g)$ and $R(g)$).
The equation (\ref{e3}) becomes meaningful for a proper choice of $g(x)$ ensuring the presence of a constant term in the right-hand
 side which connects the energy in the left-hand side. The remaining part of equation (\ref{e3}) gives the potential. Corresponding bound
state wave functions involving the special function $F(g)$ are obtained with the help of equations (\ref{e18}) and (\ref{e7}), as
\begin{equation}
\psi(x) \sim g'(x)^{-1/2} ~ e^{^{\frac{1}{2} \int Q(g) dg}}~ F(g(x)).\label{e24}
\end{equation}
Here, we choose the special function to be the exceptional $X_1$ Laguerre polynomial viz, $F(g) \propto \hat{L}_n^{(a)}(x)$.
For real $a>0$ and $n=1,2,3...$, these polynomials $\hat{L}_n^{(a)}(x)$ satisfy the differential equation \cite{UKM10a}
\begin{equation}
x \frac{d^2y}{dx^2} - \frac{(x-a)(x+a+1)}{x+a} \frac{dy}{dx} + \left(\frac{x-a}{x+a} +n-1 \right) y = 0.
\end{equation}
The polynomial $\hat{L}_n^{(a)}(x)$ has one zero in $(-\infty,-a)$,
remaining $n-1$ zeros lie in $(0,\infty)$ .
Moreover, these polynomials are orthonormal \cite{UKM09} with respect to the rational weight $\widehat{W} = \frac{e^{-x} x^a}{(x+a)^2}$
\begin{equation}
\int_0^\infty \frac{e^{-x} x^a}{(x+a)^2} \hat{L}_n^{(a)}(x) \hat{L}_m^{(a)}(x) dx = \frac{(a+n)\Gamma(a+n-1)}{(n-1)!}\delta_{nm}.\label{e28}
\end{equation}
The expressions for $Q(g)$ and $R(g)$, corresponding to the choice $F(g) = \hat{L}_n^{(a)}(x)$, are given by
\begin{equation}
Q(g) = - \frac{(g-a)(g+a+1)}{g(g+a)}, ~~~~ R(g) = \frac{g-a}{g(g+a)} + \frac{n-1}{g}.\label{e27}
\end{equation}
Using them in equation (\ref{e3}), we have the expression for $E-V(x)$ as
\begin{equation}
E-V(x) = \frac{g'''}{2g'} -\frac{3}{4} \left(\frac{g''}{g'}\right)^2 + \frac{(2 n a + a^2 - a +2)g'^2}{2 a g} - \frac{g'^2}{a(g+a)}-\frac{(a^2-1) g'^2}{4 g^2} - \frac{2 g'^2}{(g+a)^2} - \frac{g'^2}{4}\label{e4}
\end{equation}
At this point we choose $g'^2/g = \mbox{constant} = k^2, k \in \mathbb{R}-\{0\}$, which is satisfied by
\begin{equation}
g(x) = \frac{1}{4} (k x +d)^2,\label{e29}
\end{equation}
where $d$ is an arbitrary constant of integration. Here two cases may arise, namely, $d=0$ and $d\ne0$. Without loss of generality we can choose, for the moment, $d=0$.
For this choice, substituting $g(x)$ in equation (\ref{e4}) and separating out the potential and the energy, we have
\begin{equation}\begin{array}{ll}
V(x) = \frac{k^4 }{16} x ^2 + \left(a^2-\frac{1}{4}\right) \frac{ 1}{ x^2} + \frac{4 k^2}{k^2 x^2 + 4 a} - \frac{32 a k^2}{(k^2 x^2 + 4 a)^2},
\\\\
E_n = \frac{k^2( 2n +a-1)}{2}.
\end{array}\label{e5}
\end{equation}
The potential $V(x)$ is singularity free in the interval $0<x<\infty$. The same potential has earlier been reported in ref.\cite{Qu08}.
It has been shown that the potential $V(x)$ is the extension of the standard radial oscillator by addition of last two rational terms.
Such terms do not change the behavior of the potential for large
values of $x$, while small values of $x$ produce some drastic effect on the minima of the potential. The normalized wave functions corresponding to the potential
can be determined, in terms of Laguerre $X_1$ EOPs, using equations (\ref{e24}), (\ref{e28}) and (\ref{e27}), as
\begin{equation}
\psi_n(x) = \left( \frac{(n-1)!~ k^{2 a +2}}{2^{2a-3} (a +n) \Gamma(a+n-1)}\right)^{\frac{1}{2}} ~\frac{ x^{a+\frac{1}{2}}}{k^2 x^2 + 4 a} ~~e^{-\frac{k^2 x^2}{8}} ~~ \hat{{L}}_n^{(a)}
\left(\frac{k^2 x^2}{4}\right), ~~ n=1,2,3...\label{e34}
\end{equation}
It is worth mentioning here that the choice $d=0$ in equation (\ref{e29}) always gives rise to Hermitian potential.
Nonzero real values of $d$ do not make any significant difference
in the potential and its solutions.
The non-Hermiticity can be invoked into the potential only if $d$ is purely imaginary. We set $d= i \epsilon$, $\epsilon \in \mathbb{R}-\{0\}$, and $g(x) = \frac{1}{4}(k x + i \epsilon)^2$
for which the potential reduces to
\begin{equation}
\widetilde{V}(x) = \frac{k^2(k x + i \epsilon)^2}{16} + \frac{k^2 (a^2-\frac{1}{4})}{(k x + i \epsilon)^2} + \frac{4 k^2}{(k x + i \epsilon)^2 + 4 a}
- \frac{32 a k^2}{ [(k x + i \epsilon  )^2 + 4 a]^2}\label{e33}
\end{equation}
\begin{figure}[htb]
\centering

\subfigure[]{
   \includegraphics[width=3.2in] {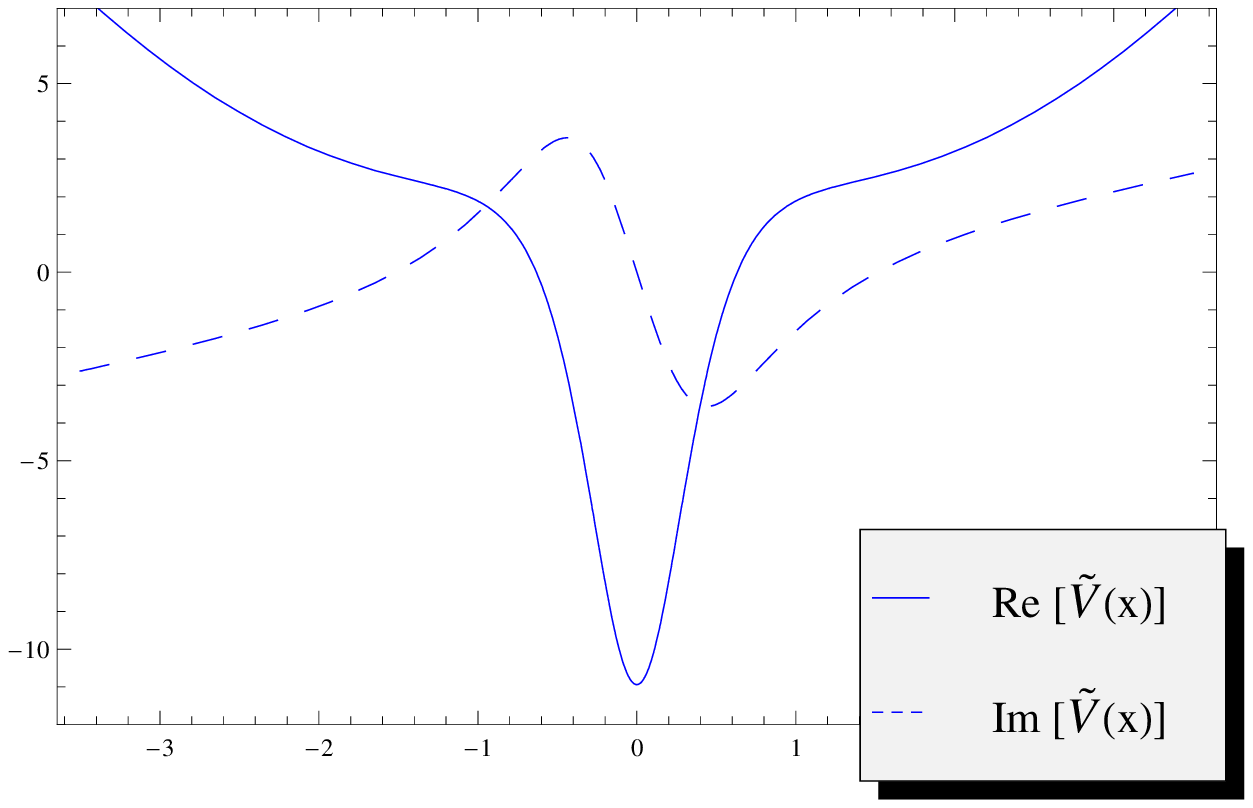}
   \label{f1a}
}
 \subfigure[]{
   \includegraphics[width =3.2in] {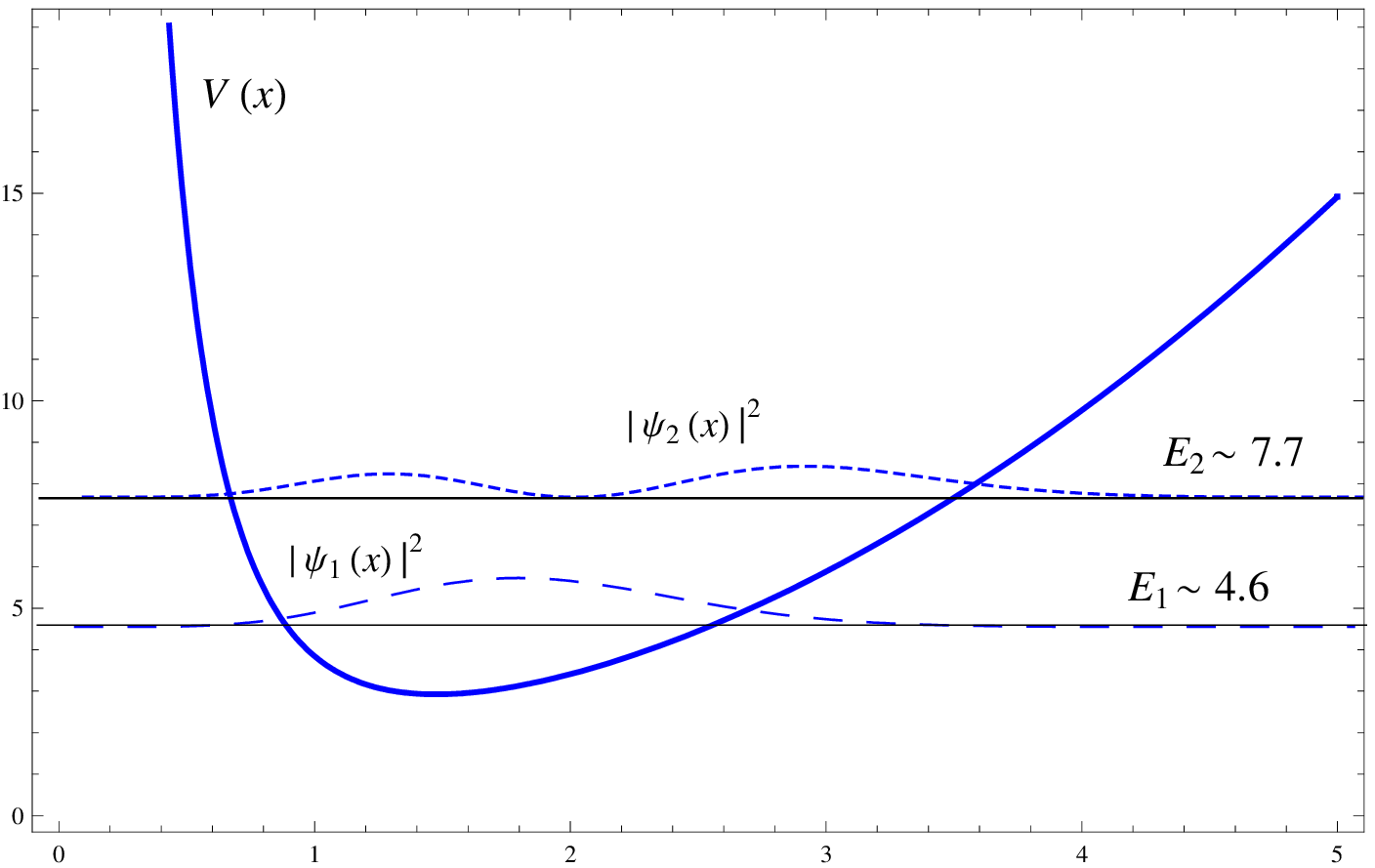}
   \label{f1b}
}
\caption{(a) Plot of the real (solid line) and imaginary (dashed line) parts of the quasi-Hermitian Potential $\widetilde{V}(x)$ associated with
 $X_1$ Laguerre EOPs. ~~(b). Plot of the corresponding equivalent Hermitian potential $V(x)$ (thick line) and square of the absolute value of its lowest
two wave functions. Here we have considered $\epsilon = 1.2,
a = 2, k = 1.75$.}
\end{figure}
The above non-Hermitian potential is free from singularity through out the whole real $x$ axis. Since the energy $E_n$ has no dependence on $d$,
the non-Hermitian Potential $\widetilde{V}(x)$ also shares the same real energy spectrum of $V(x)$. This requires further explanation. In the following, we show that the potential $\widetilde{V}(x)$ is actually quasi-Hermitian. For this we define the operator
\begin{equation}
\rho = e^{\frac{\epsilon}{k} p} , ~~p = -i \frac{d}{dx}\label{e9}
\end{equation}
which has the following properties
\begin{equation}
\rho x \rho^{-1} = x - \frac{i \epsilon}{k},~~ \rho p \rho^{-1} = p, ~~ \rho f(x) \rho^{-1} = f\left(x - \frac{i \epsilon}{k}\right). \label{e6}
\end{equation}
In other words, the operator $\rho$ has an effect of shifting the coordinate $x$ to $x - \frac{i \epsilon}{k}$.
For the proof of the results (\ref{e6}), readers are advised to follow the reference \cite{Ah01}.
For this operator we have the following similarity transformation
\begin{equation}
\rho \widetilde{V}(x) \rho^{-1} = \widetilde{V} \left(x - \frac{i \epsilon}{k}\right) = V(x).
\end{equation}
This ensures that the non-Hermitian Hamiltonian corresponding to the potential $\widetilde{V}(x)$ is quasi-Hermitian.
 The equivalent Hermitian potential $V(x)$, which corresponds to $d=0$, is given in equation (\ref{e5}).
It is very easy to show that the positive definite operator $\eta = \rho^2$ satisfies
$\eta \widetilde{V}(x) \eta^{-1} = \widetilde{V}^\dag(x)$ ensuring the potential to be pseudo-Hermitian. The potential $\widetilde{V}(x)$ also satisfies
$\widetilde{V}^*(-x) = \widetilde{V}(x)$ and hence is $\mathcal{PT}$-symmetric.
The wave functions of the potential $\widetilde{V}(x)$ can be determined by $\widetilde{\psi}_n (x) = \rho^{-1} \psi_n (x) = \psi_n \left(x + \frac{i \epsilon}{k}\right)$.\\
In figure \ref{f1a}, we have shown the real and imaginary parts of the potential $\widetilde{V}(x)$ given in (\ref{e33}), while figure \ref{f1b} shows
its equivalent Hermitian
analogue $V(x)$ given in (\ref{e5}). Using first two members of exceptional $X_1$ Laguerre polynomials
$\hat{L}_1^{(a)}(x) = -x-a-1, \hat{L}_2^{(a)}(x) = x^2 -a(a+2)$,
we have also plotted in figure \ref{f1b}
the absolute value of
first two wave functions given in (\ref{e34}).

Next we choose $F(g)$ to be Jacobi-type $X_1$ EOP, $\hat{P}_n^{(a,b)}$, which is defined for real $a,b >-1$, $a\ne b$ and $n=1,2,3..$. In this case the expression for $Q(g)$ and $R(g)$ are given by \cite{UKM09}
{\small \begin{equation}
Q(g) =-\frac{(a+b+2)g + a -b}{1-g^2} - \frac{2(b-a)}{(b-a)g - b -a}, ~R(g) = -\frac{(b-a)g -(n+a+b)(n-1)}{1-g^2} - \frac{(a-b)^2}{(b-a)g -b-a}
\end{equation}}
Using these expressions in (\ref{e3}) and choosing $g'^2/(1-g^2) = \mbox{constant} = k^2 (k \ne 0)$, we have
\begin{equation}
g(x) = \sin (k x + d).\label{e38}
\end{equation}
Like the exceptional Laguerre polynomials, the choice $d=0$ gives rise to the potential, energies and corresponding bound state wave functions, as
\begin{equation}\begin{array}{ll}
V(x) =  \frac{k^2 (2 a^2 +2 b^2 -1)}{4} \sec^2 k x - \frac{k^2(b^2-a^2)}{2} \sec k x \tan k x + \frac{2 k^2(a+b)}{a+b - (b-a) \sin k x} - \frac{8 k^2 a b}{\left[a+b - (b-a) \sin k x\right]^2},\\ \\
E_n = \frac{k^2}{4} (2 n + a + b -1)^2,
\end{array}\label{e8}
\end{equation}
and
\begin{equation}
\psi_n(x) \approx ~ \frac{(1- \sin k x)^{\frac{a}{2}+\frac{1}{4}} (1+ \sin kx)^{\frac{b}{2}+
\frac{1}{4}}}{a+b- (b-a) \sin k x}~ \hat{P}_n^{(a,b)}(\sin k x), ~~~~ ~~~n= 1, 2, 3...\label{e39}
\end{equation}
respectively. The above periodic potential $V(x)$, which is free from singularity in the interval $-\frac{\pi}{2 k} <x < \frac{\pi}{2 k}$,
can be interpreted \cite{Qu08} as the rational extension of the standard trigonometric scarf potential which is
associated with classical Jacobi polynomials. The wave functions in equation (\ref{e39}) are regular \cite{Le} iff $a,b>-1/2$.\\
\begin{figure}[htb]
\centering
\subfigure[]{
   \includegraphics[width =3.3in] {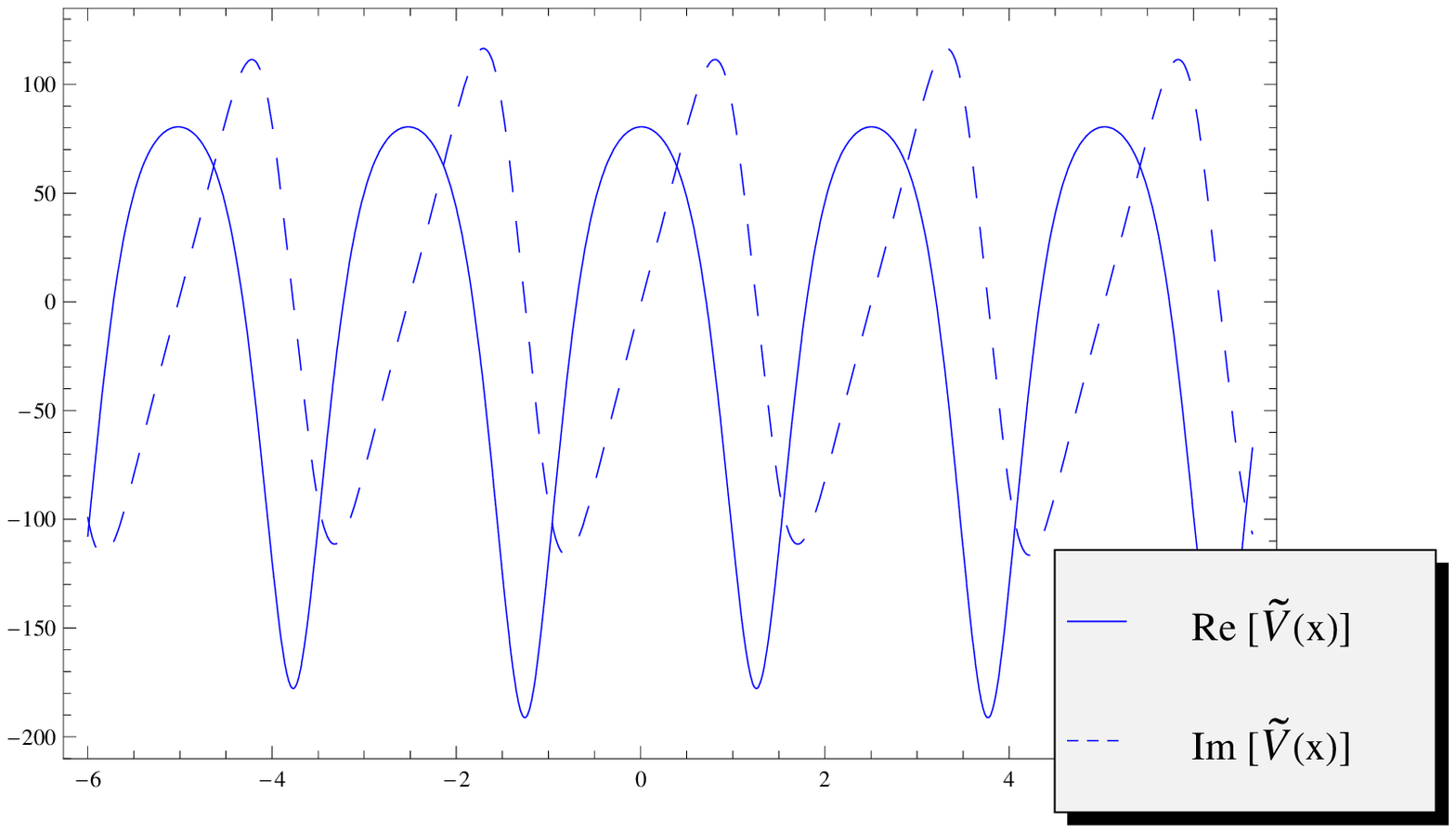}
   \label{f2a}
}
 \subfigure[]{
   \includegraphics[width =3in] {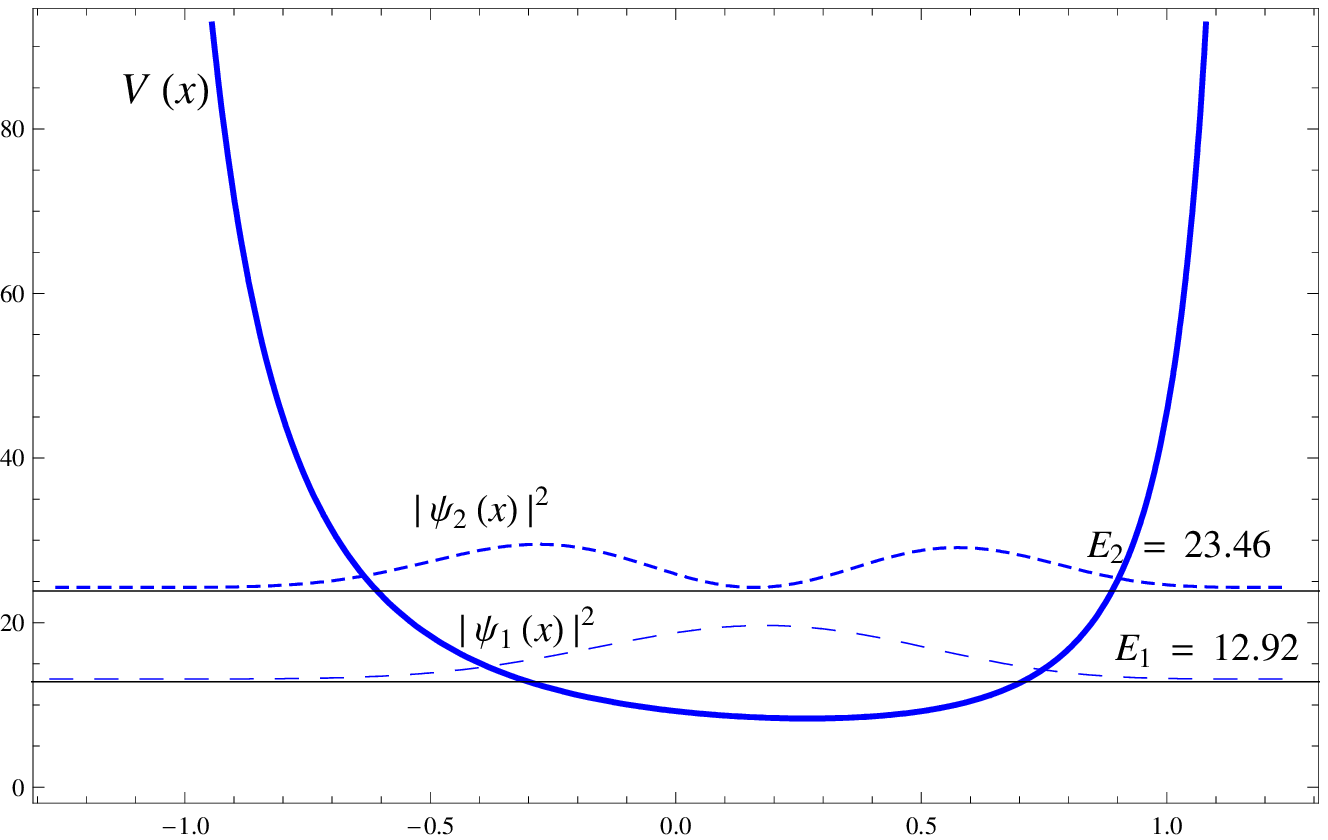}
   \label{f2b}
}

\label{myfigure}
\caption{ Plot of the real (solid line) and imaginary (dashed line) parts of the quasi-Hermitian Potential $\widetilde{V}(x)$
 associated with $X_1$ Jacobi EOPs. ~~(b).
Plot of the corresponding equivalent Hermitian potential $V(x)$ (thick line) and square of the absolute value of its lowest two wave functions. Here we have considered $a = 1.75,
b = 3,
k = 1.25,
\epsilon = 1$.}
\end{figure}
Here, the non-Hermitian potential corresponding to the choice $d= i \epsilon$ is obtained as
\begin{equation}\begin{array}{ll}
\widetilde{V}(x) = \frac{k^2 (2 a^2 +2 b^2 -1)}{4} \sec^2 (k x + i \epsilon) - \frac{k^2(b^2-a^2)}{2} \sec (k x + i \epsilon) \tan (k x + i \epsilon)\\ \\
 ~~~~~~~~~~~~~~+ \frac{2 k^2(a+b)}{a+b - (b-a) \sin (k x + i \epsilon)} +
 \frac{2 k^2[ (a-b)^2 - 4 a b]}{\left[a+b - (b-a) \sin (k x + i \epsilon) \right]^2}
\end{array}
\end{equation}
This potential $\widetilde{V}(x)$, which is defined on whole real line, also shares the same real eigenvalues of the potential given in (\ref{e8}). Like the rationally extended radial oscillator the above non-Hermitian potential is also quasi-Hermitian under the the
operator $\rho$ defined in (\ref{e9}). The corresponding equivalent analogue is the one given in equation (\ref{e8}) which corresponds to the choice $d=0$. The potential $\widetilde{V}(x)$ also fulfills the requirement of $\mathcal{PT}$-symmetry i.e. $\widetilde{V}^*(-x) = \widetilde{V}(x)$, only if $a=b$. However, if we consider the other solution $g(x) = \cos(k x+d)$ of $g'^2/(1-g^2) = k^2$, the corresponding potential $\widetilde{V}(x)$ becomes $\mathcal{PT}$-symmetric for all real values of $a$ and $b$.  The wave functions of $\widetilde{V}(x)$ can be determined by operating $\rho^{-1}$ on $\psi_n$ given in (\ref{e39}). Here, we have not considered the complex values of
$a,b$ because this will give rise to the exceptional Jacobi polynomials with complex indices and complex arguments. The orthogonality properties for such
 complex polynomials may depend on the interplay between integration contour and parameter values.\\
In figure \ref{f2a}, we have plotted the real and imaginary parts of the potential $\widetilde{V}(x)$. The corresponding equivalent
Hermitian analogue $V(x)$ and square of its first two wave functions are plotted in figure \ref{f2b}. We have used the expression of first two members of Jacobi type $X_1$ EOPs, $\widehat{P}^{(a,b)}_1 = -\frac{x}{2} - \frac{2+a+b}{2(a-b)}$ and $\widehat{P}^{(a,b)}_2 = -\frac{a+b+2}{4} x^2 - \frac{a^2+b^2+2(a+b)}{2(a-b)}x - \frac{a+b+2}{4}$ to plot the square of the wave functions.\\

\section{Summary}
In summary, we have generated some exactly solvable non-Hermitian Hamiltonians
 whose bound state wave functions are associated with Laguerre and Jacobi-type $X_1$ exceptional orthogonal polynomials.
The Hamiltonians are shown, with the help of imaginary shift of coordinate, to be both quasi and pseudo-Hermitian. The imaginary
shift of the coordinate enables us to make the potentials singularity free throughout the whole real axis.
The obtained potentials enlarge the class of analytically solvable non-Hermitian potentials. In addition, the non-Hermitian rationally extended trigonometric scarf potential
 might has potential application in
$\mathcal{PT}$-symmetric optical lattice \cite{Ma+10,Ru10}. It is to be noted here that the other choices of $g(x)$ in the expression $E-V(x)$ associated with Laguerre and Jacobi EOPs give rise to the several other exactly solvable Hermitian as well as quasi-Hermitian extended potentials. But in all these cases we have to redefine the parameters carefully so that $n$ dependent term appears only in the constant energy.

We emphasize that analogous
study \cite{MR12} can be made to the case of solvable Hamiltonians associated with exceptional orthogonal polynomials of higher co-dimension and multi-indexed polynomials.\\

\section*{Acknowledgment} I thank Barnana Roy for helpful discussion.


\begin{thebibliography}{100}
\bibitem{UKM10a} D. Gomez-Ullate, N. Kamran and R. Milson, J. Approx. Theory 162(2010)987.
\bibitem{UKM09} D. Gomez-Ullate, N. Kamran and R. Milson, J. Math. Anal. Appl. 359(2009)352.
\bibitem{UKM10b}  D. Gomez-Ullate, N. Kamran and R. Milson, J. Phys. A 43 (2010) 434016.
\bibitem{Qu08} C.Quesne, J. Phys. A 41 (2008) 392001.
\bibitem{Qu09} C. Quesne, SIGMA 5 (2009) 084.
\bibitem{MR09} B. Midya and B. Roy, Phys. Lett. A 373 (2009) 4117.
\bibitem{BQR09} B. Bagchi, C. Quesne and R. Roychoudhury,  Pramana J. Phys. 73 (2009) 337.
\bibitem{UKM12} D. Gomez-Ullate, N. Kamran and R. Milson, Cont. Math  563 (2012) 51.
\bibitem{OS10} S. Odake and R. Sasaki, Phys. Lett. B 684 (2010)173.
\bibitem{OS09a} S. Odake and R. Sasaki, Phys. Lett. B 679 (2009) 414.
\bibitem{UKM12b} D. Gomez-Ullate, N. Kamran and R. Milson, J. Math. Anal. Appl. 387 (2012) 410.
\bibitem{OS11b} S. Odake and R. Sasaki, Phys. Lett. B 702 (2011) 164.
\bibitem{STZ10} R. Sasaki, S. Tsujimoto, and A. Zhedanov, J. Phys. A 43 (2010) 315204.
\bibitem{Qu11} C. Quesne, Mod. Phys. Lett. A 26 (2011) 1843.
\bibitem{Qu11a} C. Quesne, Int. J. Mod. Phys. A 26 (2011) 5337.
\bibitem{Gr11b} Y. Grandati, J. Math. Phys. 52 (2011) 103505.
\bibitem{Gr11a} Y. Grandati, Ann. Phys. 326 (2011) 2074.
\bibitem{HOS11} C-L. Ho, S. Odake,and R. Sasaki, SIGMA 7 (2011) 107.
\bibitem{OS11a} S. Odake and R. Sasaki, Prog. Theor. Phys. 125 (2011) 851.
\bibitem{OS09b} S. Odake and R. Sasaki, Phys. Lett. B 682 (2009) 130.
\bibitem{Ho11b} C-L, Ho, Annals Phys. 326 (2011)797.
\bibitem{Ho11a} C-L, Ho, Prog. Theor. Phys. 126 (2011) 185.
\bibitem{DR11}  D. Dutta and P. Roy, J. Math. Phys 52 (2011) 032104.
\bibitem{RP+12} S. Sree Ranjani1, P.K. Panigrahi, A. Khare, A.K. Kapoor and A. Gangopadhyaya, J. Phys. A 45 (2012) 055210.
\bibitem{MRT12} B. Midya, B. Roy and T. Tanaka, J.Phys. A 45 (2012) 205303.
\bibitem{Ta10} T. Tanaka, J. Math. Phys. 51 (2010) 032101.
\bibitem{Be07} C.M. Bender, Contm. Phys. 46 (2005) 277; Rept. Prog. Phys.70 (2007) 947.
\bibitem{SGH92}  F.G. Scholtz, H.B. Geyer, and F.J.W. Hahne, Ann. Phys. 213 (1992) 74.
\bibitem{Mo10} A. Mostafazadeh, Int. J. Geom. Meth. Mod. Phys. 7 (2010) 1191.
\bibitem{BB98} C. M. Bender and S. Boettcher, Phys. Rev. Lett.  80 (1999) 5243.
\bibitem{Mo02a} A. Mostafazadeh, J.Math. Phys. 43 (2002) 205.
\bibitem{Mo02b} A. Mostafazadeh, J. Math. Phys. 43 (2002) 2814.
\bibitem{Mo02c} A. Mostafazadeh, J. Math. Phys. 43 (2002) 3944.
\bibitem{DG09} A. Das and L. Greenwood, Phys. Lett. B 678 (2009) 504.
\bibitem{MB04}  A. Mostafazadeh and A. Batal, J. Phys. A 37 (2004) 11645.
\bibitem{KS04}  R. Kretschmer and L. Szymanowski, Phys. Lett. A 325 (2004) 112.
\bibitem{ZG06} M. Znojil and H.B. Geyer, Phys. Lett. B 640 (2006) 52.
\bibitem{Ah01b} Z. Ahmed, Phys. Lett. A 282 (2001) 343.
\bibitem{BR00} B. Bagchi and R. Roychoudhury, J. Phys. A: Math. Gen. 33 (2000) L1
\bibitem{BQ02} B. Bagchi and C. Quesne, Phys. Lett. A 300 (2002) 18.
\bibitem{Le} G. Levai, Czech. J. Phys. 56 (2006) 953.
\bibitem{Zn03} M. Znojil, J. Phys. A: Math. Gen. 36 (2003) 7639.
\bibitem{LZ00} G. Levai and M. Znojil, J. Phys. A 33 (2000) 7165.
\bibitem{LZ01} G. Levai and M. Znojil, Mod. Phys. Lett. A 30 (2001) 1973.
\bibitem{Ah01} Z. Ahmed, Phys. Lett. A 290 (2001) 19.
\bibitem{Ah02} Z. Ahmed, Phys. Lett. A 294 (2003) 287.
\bibitem{BS62} A. Bhattacharjee and E. C. G. Sudarshan , Nuovo Cimento 25 (1962) 864.
\bibitem{Le89} G. Levai, J.Phys.A 22 (1989) 689
\bibitem{Ma+10} K. G. Makris et al., Phys. Rev. A 81 (2010) 063807.
\bibitem{Ru10} C. E. Ruter et al., Nature Phys. 6 (2010) 192.
\bibitem{MR12} B. Midya and B. Roy, in preparation
\end{thebibliography}
\end{document}